# Comprehensive Review of Doppler Shift Localization Methods: Advances, Limitations, and Research Opportunities

R. Szczepanik
*Military University of Technology, Warsaw, Poland*

ABSTRACT: Reliable geolocation of non-cooperative emitters in environments where Global Navigation Satellite Systems (GNSS) are unavailable or degraded is a key enabler for spectrum regulation, emergency response, autonomous mobility, and Integrated Sensing and Communication (ISAC) services in 5G/6G systems. Doppler-based techniques – from single-receiver Signal Doppler Frequency (SDF) fixes through multi-node Frequency Difference of Arrival (FDOA) and Direct Position Determination (DPD) to derivative-enhanced and learning-assisted hybrids – exploit radial-velocity-induced frequency shifts as a passive, high-resolution localization cue accessible with commodity software-defined radios, millimeter-wave access points, or acoustic sensors. This review consolidates over a decade of research across radio, acoustic, and satellite domains. It introduces a unifying taxonomy that divides the field into five technique families, outlining their evolution, measurement models, and estimator archetypes. It then compares algebraic, Bayesian, convex, and neural inference frameworks under realistic impairments such as oscillator drift, multipath, and asynchronous clocks, highlighting conditions where derivative Doppler metrics tighten the Cramér–Rao bound with minimal hardware cost. Environment-specific deployments are examined, from urban canyons and GNSS-denied tunnels to underwater, radar, UAV-swarm, and multi-orbit satellite scenarios, with prototype accuracies reaching meter scale using low-size, weight, and power payloads. Finally, the survey distils design recommendations for mobile and tactical operations and identifies open research challenges in frequency-reference integrity, multipath-aware modelling, edge-constrained computation, and trajectory-aware sensing. The findings indicate that Doppler observables, enriched by derivatives and fused with delay or angle cues, are poised to become a core positioning resource for autonomy-driven future networks.

## 1 INTRODUCTION

Accurately inferring the position of unseen signal sources has become a strategic capability for today's hyper-connected world. Telecommunications networks rely on geolocation to police spectrum use, rescue services to find survivors whose devices have no GNSS coverage, and industrial systems to coordinate fleets of drones or autonomous vehicles in real time. At the opposite end of the scale, space-borne platforms triangulate emergency beacons across oceans while underwater modems guide swarms of robots through dimly lit subsea corridors. In every case, a common technical thread emerges: localization must be delivered without relying on the co-operation of the emitter or the continuous availability of satellite timing, and it must withstand environments where multipath, interference, and adversarial behaviour are the norm.

### 1.1 *Doppler information as a localization primitive*

Among the many physical observables that can anchor a position solution – Time of Arrival (TOA), Angle of Arrival (AOA), Received Signal Strength (RSS) – the shift in carrier frequency induced by relative motion stands out for three reasons.

I. **Passive observability** – Doppler shifts arise naturally whenever transmitter and receiver exhibit relative radial motion; they can therefore be harvested by a single listening node without synchronization or handshake.

II. **Fine-grained kinematic content** – The instantaneous frequency, its first derivative (Doppler rate), and higher-order terms embed geometry and dynamics, creating a rich measurement space that complements delay-based cues.

III. **Compatibility with commodity hardware** – Modern software-defined radios (SDR), millimeter-wave access points, and even low-cost acoustic sensors can resolve sub-hertz frequency changes, enabling Doppler-centric localization to piggy-back on existing communication infrastructure.

These properties explain why Doppler shift processing has migrated from its traditional radar stronghold into cellular, Wi-Fi, IoT, and satellite constellations, and why it is attracting renewed attention in the emerging paradigm of 6G Integrated Sensing and Communication (ISAC).

### 1.2 *From single-receiver fixes to derivative-rich inference*

Classic Signal Doppler Frequency (SDF) [1] technique shows that a lone mobile receiver can localize a non-cooperative emitter to meter-level accuracy, provided its trajectory samples a sufficiently diverse set of radial velocities. Multi-receiver Frequency Difference of Arrival (FDOA) systems extend the idea to static ground stations or airborne networks, forming iso-Doppler hyperboloids whose intersections reveal the target coordinates. Direct Position Determination (DPD) engines go one step further by ingesting raw complex samples and fusing delay, angle, and Doppler signatures in a single optimization. This sharpens the cost surface and lifts identifiability in dense or low SNR scenes.

Recent work pushes the envelope in directions:

- Derivative and differential metrics – time derivatives of Doppler or differences between Doppler traces observed at multiple instants break classical geometric blind spots and tighten Cramér–Rao bounds (CRB) with negligible hardware change.
- Hybrid cue fusion – Combining Doppler with angle, delay, or range measurements mitigates the weaknesses of any single observable, producing resilient estimators for deep-urban canyons, tunnels, and underwater channels.

Learning-assisted processing – Siamese nets for channel charting, Long Short-Term Memory (LSTM) for micro-Doppler recognition, and physics-informed graph models are beginning to replace hand-crafted pipelines, promising robust operation with unsynchronized clocks, sparse arrays, or severe multipath.

### 1.3 *Aim and scope of this review*

The present survey distils a decade of research across radio, acoustic, and hybrid sensing domains. Its specific goals are to:

- Systematize the field by proposing a unifying taxonomy that covers single-sensor SDF, multi-sensor FDOA, Doppler-enabled DPD, hybrid fusions, and cutting-edge derivative methods.
- Compare estimation frameworks – from algebraic least-squares and maximum-likelihood solvers to Kalman, particle-filter, and convex-relaxation approaches – under realistic impairments such as oscillator drift, asynchronous clocks, and non-line-of-sight (NLOS) propagation.
- Map application landscapes ranging from smart-city spectrum policing, vehicular navigation, and drone swarms to underwater IoUT, through-obstacle radar, and satellite search-and-rescue.
- Expose open challenges and research trends, including frequency-reference integrity, multipath-aware modelling, edge-compute constraints, trajectory-aware sensing, and AI-enabled calibration.

The review is organized as follows. **Section II** presents a comprehensive classification of Doppler-based localization techniques, dividing the field into five key families: single-receiver SDF, multi-receiver FDOA, Doppler-aided DPD, hybrid approaches combining Doppler with delay or angle cues, and cutting-edge methods utilizing Doppler derivatives and machine learning. It also outlines the historical development and technical foundations of each category.

**Section III** formalizes the mathematical models linking Doppler observables (and their derivatives) to emitter–receiver kinematics and surveys a range of estimation algorithms – from maximum likelihood and Kalman filters to convex optimization and data-driven methods – under practical impairments such as oscillator drift, multipath propagation, and clock asynchrony.

**Section IV** explores how these techniques are adapted to specific environments, including urban canyons, tunnels, underwater domains, radar scenes, UAV swarms, and satellite networks, highlighting how Doppler processing is tailored to propagation conditions and hardware constraints.

**Section V** reviews experimental platforms and real-world deployments, detailing hardware choices (e.g., SDRs, atomic clocks), system architectures (e.g., UAV-based sensors, GNSS-denied navigation), and field results demonstrating meter-scale accuracy in diverse scenarios.

**Section VI** discusses key open challenges – such as frequency-reference instability, multipath bias, estimator complexity, and edge-computing limitations – and analyzes their impact on localization accuracy and system scalability.

**Section VII** outlines future research directions, emphasizing the integration of machine learning, trajectory-aware sensing, multi-agent cooperation, and ISAC-compatible waveform design as critical enablers for robust Doppler-based positioning in 6G and autonomy-driven networks.

By weaving together theory, algorithms, hardware realizations, and use-case-driven insights, I hope to provide newcomers and seasoned practitioners with a coherent picture of Doppler-based localization and where it is heading next.

## 2 CLASSIFICATION OF LOCALIZATION TECHNIQUES USING THE DOPPLER EFFECT

### 2.1 *Single-Receiver Techniques: Signal Doppler Frequency*

The SDF method exploits the functional relationship between the instantaneous Doppler shift observed by a moving receiver and the relative geometry of the transmitter–receiver pair. Because only one sensor is required, SDF offers a uniquely lightweight solution for rapid deployment in scenarios where time synchronization, network infrastructure, or cooperative transmitters are unavailable.

Early analytical work formalized the forward model that links the Doppler trace to emitter position. A closed-form wave-equation solution for a moving source in free space was derived in [2], establishing how trajectory curvature and velocity manifest in the received frequency. Practical inversion procedures were introduced soon after: a 2D/3D algorithm delivering meter-level accuracy from an airborne platform is detailed in [3], while the extension to complex PSK waveforms via higher-order spectral peaks appears in [4] and [5]. A broader survey of SDF applications, from electronic reconnaissance to disaster-relief search, is presented in [6].

**Algorithmic advances.**

Several studies focus on increasing accuracy and computational robustness without sacrificing the one-receiver premise. A twin-stage overlapping window estimator that blends IQ-domain pre-filtering with Gaussian smoothing reaches sub-5 m error from an Unmanned Aerial Vehicle (UAV) in real flights [7]. Frequency-domain marker extraction enables simultaneous localization of multiple co-channel emitters with relative errors below 1 % after a few minutes of flight [8]. Studies of multipath and NLOS show that weighted or maximum-weight Doppler estimators can cut the error by an order of magnitude in dense urban layouts [9]. At the theoretical end, Doppler-derivative augmentations shrink blind geometries and bring the Cramér-Rao bound within reach [10].

**Hardware implementations and validation.**

SDR prototypes dominate recent experimental work. A real-time LabVIEW & MATLAB implementation on a USRP B200mini achieves 2 – 5 m accuracy in suburban drive tests [11]. Controlled-range emulation using the same hardware quantifies how internal oscillators inflate the location error and confirms the benefit of chip-scale atomic clocks (CSAC) [12]. A systematic study across six low-cost SDRs demonstrates that external rubidium disciplining can reduce 10 km range errors from kilometer to single-meter scale [13]. Field trials of the fully autonomous ASLER system, mounted on a multirotor, report an 8.7 m mean error at 250 m stand-off while producing live map overlays through offline cartographic data [14] and GNSS fusion [15].

**Operational extensions.**

Research has broadened SDF beyond classical single-target, free-space assumptions. An acoustic laboratory surrogate proves the concept for VHF/UHF radios when physical motion is impractical [16]. Automatic VTOL landing guidance in GNSS-denied sites attains sub-meter touchdown precision using a sparse network of narrow-band beacons [17]. Cooperative-swarm studies illustrate how multiple UAVs, each running its own SDF processor, can combine estimates to maintain track continuity of a maneuvering emitter in harsh line-of-sight (LOS)/NLOS transitions [1].

**Current challenges and outlook.**

Key limitations remain. Frequency instability – induced bias, especially in commercial SDR front-ends, still drives the error budget; temperature-adapted oscillators or over-the-air calibration are active topics [13]. Multipath rejection in cluttered indoor or canyon environments, although mitigated by weighting schemes [9], necessitates joint time–frequency–angle filtering or machine-learning post-processing. Finally, latency constraints for fast-mover tracking motivate incremental estimators that fuse Doppler with bearing or TOA in real time – an avenue being pursued via pseudo-linear algebraic solvers [18] and hybrid sequential TOA/DFS filters [19].

Overall, the SDF literature demonstrates that a single, mobile receiver can localize emitters with meter-scale accuracy under realistic dynamics, provided that frequency estimation, motion sensing, and oscillator control are carefully engineered. Continued convergence of SDR platforms, UAV autonomy, and robust signal processing to cement SDF as a practical backbone for rapid deployment positioning, spectrum policing, and GNSS denied navigation.

### 2.2 *Frequency Difference of Arrival FDOA*

FDOA methods estimate the position of a non-cooperative emitter from the differences between Doppler shifts measured at two or more spatially separated receivers. Unlike single-receiver SDF, FDOA creates iso-Doppler hyperboloids whose intersection pinpoints the target. A relative motion – either of the sensors or transmitter – is mandatory to generate a non-zero frequency gradient.

**Modelling fundamentals.**

The canonical formulation, detailed exhaustively in the electronic-reconnaissance monograph [20], expresses each differential measurement as the projection of the relative velocity vector onto the difference of unit LOS vectors. The resulting surfaces are hyperbolic cylinders in 2D and double hyperboloids in 3D; closed-form Geometric Dilution of Precision (GDOP) expressions reveal severe singular geometries when the receiver baseline is nearly orthogonal to the velocity vector. Practical accuracy is further limited by Hz-level frequency-estimation noise, transmitter frequency instability, and ambiguity in wideband/LFM signals – issues quantified through CRB analysis in [20].

**Algorithmic landscape.**

Classical solutions fall into families:

- Batch maximum-likelihood (ML) solvers that search the 2D/3D space iteratively, achieving minimum variance at the cost of heavy grid scans.
- Algebraic or convex relaxations that linearize the measurement model and solve a weighted least-squares (WLS) or semidefinite programming (SDP).

A comprehensive survey of wireless positioning for moving receivers [21] maps these algorithms onto deterministic (WLS, SDP) and probabilistic (particle filter, Extended Kalman Filter (EKF), Unscented Kalman Filter (UKF)) schemes. It highlights their sensitivity to NLOS blockage and small relative velocities. Hybrid extensions – TDOA/FDOA and AOA/FDOA – are recommended to break geometrical degeneracies and improve observability when Doppler dynamics are weak [21].

**Sequential single-sensor FDOA.**

Traditional FDOA demands time-synchronized receivers; however, a recent on-the-fly technique uses one mobile sensor that records Doppler shifts at successive time instants and treats them as differential pairs [22]. By fitting iso-Doppler curves in a polar coordinate frame, the method localizes acoustic and RF beacons by computing two orders of magnitude below ML while remaining unbiased after outlier rejection. This paradigm is attractive for small unmanned platforms where size, weight, and power (SWaP) constraints preclude multi-receiver arrays.

**High-dimensional and derivative-based extensions.**

For a maneuvering objective observed by immobile arrays, closed-form and SDP solvers directly exploit the absolute Doppler shifts and their temporal derivatives to recover full 3D kinematics [23]. The inclusion of Doppler rate terms mitigates blind cones. It eliminates frequency-grid searches, bringing performance within 1 dB of the Cramér–Rao Lower Bound (CRLB) under moderate SNR and modest sensor-placement error.

**Open challenges and research directions.**

- **Frequency reference integrity**. Hz level stability is needed to unlock high localization accuracy at VHF; oven-controlled or GNSS disciplined oscillators remain mandatory for operational deployments.
- **Multipath and NLOS**. While angle Doppler filtering helps, coherent multipath can bias small $\Delta f$ measurements; robust estimators combined with model-based multipath mitigation are still under active investigation.
- **Low-Doppler regimes**. Urban macro-cells or slow UAV swarms generate Doppler spreads of only a few hertz; hybridization with TOA or SDF measurements – as advocated in [21] – offers a promising remedy.
- **Autonomous mobile FDOA**. Sequential single-receiver concepts [22] need further validation in, complete 3D trajectories and cluttered RF spectra.

In summary, FDOA provides a powerful multi-receiver alternative to SDF, achieving network-level geolocation accuracy without requiring transmitter cooperation. Advances in single-sensor schemes, convex optimization, and derivative-aided modelling continue to push the technique toward real-time, low-SWaP implementations suitable for 5G/6G, ISR, and IoT sensor-networks.

### 2.3 *DPD with Doppler Measurements*

DPD delivers source coordinates directly from raw sensor data, avoiding the intermediate extraction of AOA, TDOA, or FDOA parameters. When Doppler information is included, the localization surface gains an additional velocity-dependent dimension that sharpens the cost-function peaks and improves identifiability. The literature may be grouped into five mutually reinforcing research tracks.

**1) Computationally efficient single-source solvers**

The baseline ML grid search introduced for narrow-band emitters in [24] proved near-optimal accuracy but at prohibitive cost. Successive work focuses on accelerating the optimization.

- Iterative EM/Laplace: the Maximum Entropy Method (MEM-DPD) algorithm converts the multi-dimensional ML task into independent one-dimensional polar searches, reaching the CRB with an order-of-magnitude lower complexity [25].
- FFT-assisted ML: for Orthogonal Frequency-Division Multiplexing (OFDM) waveforms, both Multi-Step Maximum Likelihood (MSML-DPD) [26] and its simplified variant Simplified Maximum Likelihood (SML-DPD-FFT) [26] reuse a single FFT of the deconvolved signal to evaluate every candidate grid point, slashing run-time while preserving ML accuracy.

- Closed-form Doppler/Delay fusion: a Marginalized Particle Filter (MPF)-based scheme jointly updates position and velocity from raw TDOA and Doppler data, reducing the particle count – and thus CPU load – by a factor of ten without degrading accuracy at low SNR [27].

**2) High-resolution and multi-source extensions**

When several co-channel emitters are active, standard ML suffers from source-association ambiguity. High-resolution beamspace and MVDR ideas address this limitation.

- Minimum Variance Distortion less Response (MVDR-DPD) generates a fine-grained spectrum whose peaks reveal the emitter set without prior knowledge of source number [28]; a dedicated Doppler-driven variant, DPD High Resolution (HR), outperforms classical ML in low-SNR or bursty-signal regimes [29].
- Beamspace-DPD compresses array data to a reduced subspace, cutting data-rate and CPU requirements while retaining classical Multiple Signal Classification (MUSIC)-level accuracy [45].
- The MUSIC based multi-source framework of [30] combines spatial steering vectors with Doppler signatures, enabling localization of more sources than antenna elements per station.
- An importance-sampling ML reformulation eliminates grid search and offers massive parallelism on GPUs, yielding CRB-level performance for many emitters observed by moving sensors [31].
- Single-snapshot localization of several narrow-band emitters is possible when Doppler and AOA are fused at the covariance level, as demonstrated in [32].
- Qin et al. [33] propose a decoupled direct positioning algorithm tailored for strictly noncircular sources, combining Doppler shifts with AOA. The method enhances localization accuracy by introducing a modified signal model that accounts for receiver motion and the noncircular properties of signals such as BPSK. An iterative decoupling strategy is applied to the maximum likelihood search to address the computational burden in multi-source scenarios. Simulation results show that the approach outperforms both conventional DPD and two-stage methods, particularly under low-SNR conditions and in the presence of model mismatches.

**3) Robustness to multipath and large arrays**

Multipath causes spurious peaks in the DPD cost surface. Three complementary solutions emerge.

- Massive multiple-input multiple-output (MIMO) Direct Source Localization (DiSouL) separates LOS from NLOS components with compressive-sensing machinery, delivering reliable positioning in rich-scattering channels [34].
- Joint spatio-temporal suppression exploits the asymptotic orthogonality of multipath across bandwidth and aperture, achieving near-CRB bounds without explicit Doppler use but laying the groundwork for Doppler-enabled massive arrays [35].
- Off-grid sparse Bayesian inference explicitly models LOS/NLOS angle bias and updates grid offsets iteratively, outperforming super-resolution baselines on coarse grids [36].

**4) Quantization, distribution, and hardware pragmatics**

Emerging deployments tighten fronthaul and power budgets:

- One-Bit DPD preserves localization performance with mere sign-bits of the baseband samples by relying on second-order statistics [37].
- Sinc-interpolated distributed DPD passes only low-dimensional sufficient statistics among neighboring nodes, matching centralized accuracy under severe SNR and connectivity constraints [38].
- An OFDM positioning scheme that operates in fully unsynchronized networks demonstrates that Doppler-resolving DPD can coexist with legacy infrastructure, even under severe multipath [39].

**5) Specialized scenarios and outlook**

High-speed sensor trajectories demand Doppler-centric processing; MEM-DPD [25] already targets this niche, and DPD-HR for moving receivers [29] pushes resolution further. FFT-accelerated solvers [26] [25] supply the needed real-time throughput for OFDM radar-communication convergence. Future priorities include fusing Doppler derivatives, clock-offset calibration, and learning-based cost-surface shaping to extend DPD to ultra-dense 6G and ISAC networks.

Collectively, these works confirm that Doppler-aided DPD now matches or exceeds classical two-stage pipelines in accuracy while offering better scalability, multisource capability, and resilience against multipath, making it a prime candidate for next-generation passive localization systems.

### 2.4 *2.4. Hybrid Techniques that Fuse Doppler with TOA, AOA, TDOA, and other cues*

Leveraging more than one physical metric mitigates the geometric weaknesses of any single-measurement class and widens the operational envelope of Doppler-based localization systems. The papers surveyed combine frequency shifts with angles, delays, or ranges, moving progressively from single-platform 2D solvers to fully cooperative, three-dimensional, multi-agent schemes.

**Single-platform fusions of bearing and Doppler.**

For the classic problem of a lone sensor circling an emitter, three closed-form pseudolinear estimators –

PLE, its bias-compensated variant BCPLE, and a weighted instrumental-variable solution – achieve nearly Cramér–Rao performance while avoiding iterative optimization [40]. The key is a novel linearization of the Doppler equation that delivers direct 2D position fixes from simultaneous AOA and frequency observations.

**Differential-time and frequency hybrids for mobile sources.**

An algebraic two-stage WLS method blends TDOA, FDOA, and differential Doppler-rate measurements; it remains robust to receiver-position errors and reaches the CRB under moderate noise [41]. A related altitude-constrained solver incorporates TDOA, FDOA, and Differential FDOA (dFDOA) into an iterative WLS framework, offering dedicated variants that explicitly accommodate sensor kinematic uncertainties [42].

**Sequential or filter-based Doppler-TOA integration.**

The Least-squares Adaptive Sequential Doppler and Time-of-arrival (LAS-SDT) algorithm exploits a stream of Doppler shifts and TOA readings to jointly estimate position, velocity, and clock parameters of an unsynchronized, moving user; extensions that inject prior knowledge of speed or clock drift further tighten accuracy bounds [19]. In tunnels where GNSS is denied, combining LTE-V Doppler with TOA in an EKF cuts the mean localization error to 20 m without extra roadside infrastructure [43].

**Range and velocity-aided cooperative networks.**

In wireless-sensor networks, a block-coordinate-descent/online Gauss–Newton scheme reconstructs the 6-Degrees of Freedom (DOF) states of multiple rigid bodies from inter-node ranges and relative Doppler data; WLS refinement yields centimeter-level accuracy across irregular topologies [44]. For multi-rigid-body underwater Internet of Underwater Things (IoUT) scenarios, trajectory optimization guided by CRB-aware metrics fuses ToA and Doppler to steer autonomous observers along information-rich paths, markedly outclassing ToA-only baselines [45].

**Multistate sonar and radar with differential delays.**

In bistatic or multistate sonar, adding Doppler differences to time-delay measurements tightens the localization region, especially when the target is close to the transmitter–receiver baseline. Two-stage algebraic/MLE solvers demonstrate CRB-level accuracy without iteration [46], while a bias-reduced SDP formulation achieves similar gains even when the propagation speed is unknown [47]. For moving sensors tracking a moving target, ignorance of Doppler coupling can degrade accuracy by orders of magnitude; Gauss-Newton and quasi-Newton MLE refinements restore performance when full motion models are used [48].

**Application-driven hybrids.**

In Wi-Fi–based search-and-rescue, Doppler shifts from survivors' smartphones are fused with inertial sensing and frequency-offset calibration, halving search times versus RSS techniques [49]. A comprehensive tutorial dissects how such hybrid strategies can be centralized, distributed, or collaborative, reviewing particle-filter, Kalman, and direct-position approaches for 5G/6G mobile receivers and highlighting the resilience of FDOA/TDOA fusion under LOS conditions [21].

### 2.5 *Cutting-Edge Extensions: Doppler Derivatives, Differential Metrics, Micro-/Macro-Doppler, and Machine-Learning Aids*

Research at the frontier of Doppler-based positioning is moving beyond "plain" frequency-shift measurements. Recent work introduces higher-order Doppler observables, differential combinations, data–driven mapping, and micro-Doppler exploitation, often wrapped in learning-centric frameworks that dispense with classical synchronization or significant antenna assets.

**Higher-order Doppler observables.**

Ke & Ho show that the time-derivative of the Doppler shift carries complementary geometric information that breaks classical blind directions and tightens the position solution. Their closed-form and SDP solvers work whether the nominal carrier is known, attain Cramér–Rao accuracy, and remove the grid search found in earlier FDOA engines [10]. Differentiating the frequency cue foreshadows a new family of dFDOA and Differential Frequency Drift Rate (dFDR) estimators now entering the literature.

**Learning-driven channel geometry from Doppler.**

Channel charting traditionally relies on MIMO fingerprints; by contrast, a recent Siamese-network approach embeds Doppler-induced phase differences into the loss function, enabling global mapping with only four, mutually unsynchronized antennas [50]. The network aligns local channel manifolds into a coherent chart that simultaneously localizes transmitters and sketches the propagation environment, outperforming previously CSI-only charting techniques.

Complementarily, Chen et al. demonstrate that even a single-antenna user can localize and map 5G/6G multipath reflectors when its motion modulates the received spectrum. Their one-dimensional search uses Doppler on multiple NLOS paths and needs no time-base sharing between the base station and the user [51]. As the user speed grows, the position error bound shrinks, highlighting the value of deliberate mobility in Doppler-centric designs.

**Micro-Doppler signatures and deep sequences.**

While the studies above exploit macro Doppler (bulk radial velocity), radar researchers turn to micro-Doppler – fast fluctuations from propellers or limbs – for simultaneous detection, classification, and bearing-based localization of small UAVs. A bistatic radar pipeline that couples empirical-mode decomposition, Short-Time Fourier Transform (STFT), and principal-component analysis with an LSTM network yields state-of-the-art probability of detection and sub-degree angle estimates using inexpensive hardware [52]. Such fine-grained spectral fingerprints open a path toward intent recognition and target identification in ISAC scenarios.

**Towards 6G Integrated Sensing and Communications.**

Finally, a recent survey canvasses OFDM, Orthogonal Time Frequency Space (OTFS), Generalized Frequency Division Multiplexing (GFDM), and dual-purpose pilot structures that embed range, angle, and Doppler information into communication waveforms [53]. The authors distil synchronization requirements, frame structures under 5G-Advanced, and standardization gaps, providing a roadmap for AI-assisted waveform and receiver co-design where Doppler becomes a first-class citizen.

**Take-away trends.**

- **Derivative and differential Doppler metrics** offer richer observability with minimal hardware change.
- **Learning frameworks** (Siamese nets, LSTMs, self-supervised channel charting) are supplanting hand-crafted estimators, especially when synchronization, complete Channel State Information (CSI), or large antenna arrays are unavailable.
- **Micro-Doppler analysis** is emerging as a bridge between localization and object semantics, dovetailing with ISAC ambitions for 6G.

Collectively, these innovations push Doppler-aided localization from niche electronic-reconnaissance tools toward pervasive, AI-enhanced sensing layers in future wireless networks.

## 3 MEASUREMENT MODELS AND ESTIMATORS

The mathematical relation between Doppler frequency shifts (and their derivatives) and the kinematics of a transmitter–receiver pair underpins every algorithm surveyed in this review. Once linearized, the model becomes amenable to a broad spectrum of estimators – from maximum-likelihood search and expectation–maximization through Kalman and particle filtering to convex relaxations such as semidefinite programming and pseudolinear least-squares (PLE/WLS). Yet real-world deployments face three persistent hurdles: (I) **frequency instability of low-cost radios**, (II) **channel-induced errors** such as multipath or NLOS propagation, and (III) **lack of global synchronization** across sensors. The papers discussed below refine the Doppler model, propose robust estimators, and quantify the impact of practical impairments.

### 3.1 *Refining the Doppler Measurement Model*

Picard & Weiss carry out a small-error analysis (SEA) that links NLOS perturbations directly to bias and variance of Doppler-based DPD. Their closed-form expressions reveal how error magnitude and orientation depend on receiver geometry, prompting design rules such as square sensor layouts [54]. A companion study extends the analysis, validating the formulas with Monte-Carlo experiments and showing that quadrilateral arrays minimize the NLOS bias under realistic noise levels [54].

In the underwater domain, Luo et al. demonstrate that multipath reflections can be tamed by simply lengthening the observation window: the mean-square error of a DFT-based Doppler estimator decays quadratically with sample count, rendering multipath almost negligible after a few seconds of data [55].

At millimeter-wave frequencies, Fang et al. introduce a hybrid deterministic–stochastic channel model that synthesizes time-frequency signatures – including micro-Doppler streaks – from human motion inside rooms. The model furnishes a configurable test-bench for ISAC algorithms and highlights how body posture and wall materials shape the Doppler spectrum [56].

Finally, though developed for beat-tracking in music, the zero-latency PLP architecture by Meier et al. offers a transferable concept: causal time–frequency buffers coupled with look-ahead prediction to mitigate processing lag – an attractive feature for real-time Doppler tracking in acoustic or RF localization [57].

### 3.2 *Estimation Frameworks under Imperfect Conditions*

Ma, Liu & Guo tackle the doubly challenging case of asynchronous sensor networks. Their EM-Gauss-Newton routine simultaneously calibrates clock offsets and extracts the source position from TDOA measurements, cutting computation by orders of magnitude compared with brute-force grid search. Although focused on TDOA, the formulation is readily extendable to FDOA and mixed-cue scenarios, setting a template for joint synchronization-localization solvers [58].

The SEA insights mentioned earlier translate into algorithmic guidelines: weighting schemes that down-rank Doppler measurements arriving via likely NLOS paths can be embedded in Kalman or particle filters. At the same time, robust convex relaxations such as SDP absorb residual model mismatch. In underwater tracking, for instance, the bias trends quantified by Luo et al. inform adaptive window selection inside DFT-based estimators [55].

For indoor ISAC, the mmWave model of Fang et al. supports data-driven training of neural networks that regress position or gesture classes directly from raw micro-Doppler maps – an example of machine-learning estimators that bypass explicit geometric inversion [56]. The LSTM predictor of Meier et al. further illustrates how recurrent nets can capture temporal coherence in frequency-shift sequences, a capability valuable when Doppler varies rapidly with maneuvering targets [57].

### 3.3 *Impact of Hardware and Channel Impairments*

Frequency instability and oscillator drift enter the Doppler model as process noise, inflating the Cramér–Rao lower bound. EM iterations such as those in [58] can absorb quasi-static drifts, whereas extended Kalman filters equipped with state-augmented clock variables prove effective for higher-order wander.

Multipath manifests as stochastic perturbations in Doppler and delay; the SEA framework [54] quantifies this effect, while long-aperture averaging or model-based subtraction – as advocated in the underwater study [55] – provide practical mitigation.

Lastly, the asynchronous setting of [58] and the unsynchronized-antenna charting of [57] confirm that global clock alignment is no longer mandatory when estimators fuse Doppler with ancillary cues or operate in a self-supervised fashion.

In summary, current research converges on a layered strategy: (I) enrich the Doppler measurement model to capture NLOS, micro-Doppler, and clock drift; (II) deploy hybrid estimators – EM, Kalman, SDP, particle filters – tailored to the refined model; and (III) validate performance against CRB that explicitly embeds the identified impairments. The selected works [54] [55], [56], [57] [58] collectively advance each layer, laying the groundwork for robust Doppler-centric localization in the presence of real-world non-idealities.

## 4 APPLICATION-ORIENTED LANDSCAPE – OVERVIEW

Doppler-centric localization is highly context-sensitive: the signal model that holds in a glass-and-steel city block collapses in a salt-water channel, while the estimator that thrives on broadband radar pulses may fail when only sparse satellite passes are available. Urban canyons replace the direct wavefront with fast, specular reflections that bias frequency estimates and demand long-track integration or multipath-aware inference. In underwater acoustics, the propagation speed is five orders of magnitude lower than in free space; coherent observation windows stretch to seconds, bandwidth contracts, and even slight uncertainty in sound speed skews the entire geometry, prompting bias-reduced convex optimization. Confined transport corridors, such as road or rail tunnels, add yet another twist: satellite navigation drops out completely, anchor geometry becomes nearly colinear, and hybrid filters must fuse Doppler dynamics with opportunistic TOA or inertial cues to keep objects on course.

Radar and UAV scenarios illustrate the opposite extreme of plentiful bandwidth and agile motion. Here, through-obstacle tracking relies on sub-millisecond modelling of instantaneous frequency ridges, phased-array weather radars exploit space-time adaptive processing to disentangle clutter, and six-degree-of-freedom channel models reveal how rotational micro-Doppler alters both sensing and communication links. UAVs themselves switch roles between elusive targets, whose blade flashes feed LSTM classifiers, and mobile sensing platforms, where lightweight software-defined radios perform signal-Doppler fixing of ground emitters to meter-level accuracy.

At orbital altitudes, localization turns sparse again. Low-Earth-Orbit (LEO) assets follow fast ground tracks that embed a strong Doppler signature, yet require several passes to triangulate a beacon. Geostationary (GEO) satellites contribute no range-rate diversity but offer a precision frequency reference; fusing the two information types tightens the solution space, shortens rescue latency, and resolves mirror-image ambiguities. Across all these environments, a typical pattern emerges: exploit every sliver of Doppler information available, pair it with complementary observables when possible, and choose estimation techniques – Kalman, EM, particle filtering, semidefinite relaxation, or data-driven surrogates – that explicitly model the dominant biases of the medium. The following sections trace how these principles translate into concrete system designs, revealing a rich landscape in which Doppler shifts and their derivatives serve as versatile, environment-adaptive keys to positioning.

### 4.1 *Urban and NLOS scenarios*

Dense, cluttered environments fundamentally distort Doppler measurements: LOS paths are blocked, reflections create micro-multipath, and low-cost hardware drifts in frequency. The literature, therefore, centers on two complementary questions: how to extract usable Doppler (or Doppler-derived) observables when LOS is absent or intermittent, and how to build estimators that remain stable in the face of biased, asynchronous, or highly correlated data. The works reviewed below illustrate the design space – from analytic signal-processing tricks through Bayesian inference to deep-learning surrogates.

**Single SDF sensor in urban canyons.**

Early studies proved that even one moving receiver can localize emitters inside cities when Doppler traces are integrated over suitably long tracks. The 3D method originally introduced for flight rescue or reconnaissance platforms [3] was later adapted to phase-modulated signals – first in controlled field trials [4], then in live urban drives that achieved 2 – 3 m accuracy despite severe multipath [5]. Two recent upgrades tighten robustness: a two-stage overlapping window with Gaussian filtering that lifts the precision of UAV sensors below 10 m [7], and an SDF + GNSS fusion engine that overlays Doppler fixes on offline vector maps to compensate for NLOS drift (2 – 10 m error in downtown tests) [15]. A tutorial perspective on SDF-based navigation and emergency services, including GNSS-denied UAV landing aids, summarizes the operational envelope of these methods [59].

**RFID, Wi-Fi, and mmWave: Short-Range Indoor Localization.**

Indoors, tag or user motion can be exploited as a "built-in" frequency sweep. For Radio Frequency Identification (RFID), an analytic Doppler model linked the radial velocity of a handheld tag to reader-tag geometry and enabled meter-level positioning without extra infrastructure [60]. On commodity Wi-Fi, a rescue-oriented framework transforms broadband packets into narrowband surrogates, applies Doppler calibration, and fuses inertial data, halving search times in collapsed-building drills [49]. Moving up in frequency, mmWave routers offer large bandwidth and beam steering: mmTrack passively tracks multiple people to 13 cm by combining micro-Doppler slices with beam-sweep kinematics [61]. At the same time, RAPID retro-fits 802.11ay access points for human presence detection via phase micro-flutter analysis [62]. A dedicated scattering model at 28 GHz now supplies synthetic micro-Doppler data for algorithm design and repeatable benchmarking [56] [90].

**DPD under multipath and NLOS.**

DPD attempts to bypass intermediate parameter estimation, yet NLOS can bias its likelihood surface. Two companion SEA papers derive closed-form expressions for that bias and prove that a square array layout is least sensitive to NLOS angle spreads [54]. For massive MIMO, DiSouL leverages the immense spatial resolution of large arrays together with compressive sensing to separate LOS from reflections and achieve sub-meter accuracy indoors [34]. When infrastructure is limited to non-synchronized OFDM base stations, a blind cross-correlation scheme recovers a consistent time-frequency reference directly at the user node [39]. Sparse Bayesian off-grid inference further mitigates grid-quantization error and jointly classifies LOS/NLOS rays, outperforming super-resolution baselines in deep-urban trials [36]. For multi-emitter scenes, MUSIC-type cost functions fed with Doppler snapshots yield high-resolution DPD but still require centralized raw-signal fusion [30]; a recent ML importance-sampling reformulation slashes complexity and allows parallel execution on edge servers [31].

**Learning-driven models with minimal infrastructure.**

Neural approaches increasingly target scenarios where synchronization, sensor diversity, or long observation times are impractical. A Siamese-network channel-charting pipeline is trained on Doppler-induced phase differences only, reconstructing the geometry of an area with just four unsynchronized antennas and even static transmitters [50]. In cellular 5G/6G, a single-antenna user equipment can localize itself and map reflecting surfaces by exploiting Doppler signatures of multiple NLOS paths; a one-dimensional search suffices to reach the CRB at moderate SNR [51]. For wireless-sensor networks, constrained NLS fused with an EKF supports sub-meter robot navigation and human tracking without GPS by relying on infrastructure nodes that measure uplink Doppler shifts [18].

**Robust Modelling and Prediction Tools.**

Beyond classic localization, real-time beat-tracking with zero latency – initially devised for music signals – has been proposed as a proxy to predict Doppler-induced timing offsets before they occur, potentially stabilizing time-critical acoustic or RF ranging [57]. Such predictive filters could be combined with analytic bias characterization [54] and map-aided SDF engines [15] to form resilient localization stacks tailored to the deep-urban NLOS regime.

**Key Take-aways for urban/NLOS deployment:**

- **Multipath is not always an enemy**: massive-MIMO sparsity [34] [37] or micro-Doppler fingerprints [61], [62] can actively exploit reflections for added diversity.
- **Single-platform solutions mature**: UAV-borne SDF [7], [15], RFID [60], or single-antenna 5G user equipment [51] demonstrate that Doppler alone, if modelled, can close the geometry without cooperative anchors.
- **Bias awareness is critical**: analytical SEA tools [54] and Bayesian off-grid updates [36] quantify and mitigate NLOS errors instead of merely averaging them out.
- **Edge AI and channel charting reduce calibration overhead**: data-driven embeddings [50] and importance-sampling DPD [31] lessen the need for clock synchronization or exhaustive grid searches.

Collectively, these contributions show that Doppler-centric localization has progressed from a LOS-dependent niche to a viable option for dense, clutter-dominated smart-city scenarios.

### 4.2 *Underwater and sonar-based localization systems*

Underwater acoustics present a markedly different operating theatre from radio or mm-wave scenarios: the propagation speed is five orders of magnitude slower than light, bandwidth is scarce, and strong, slowly decaying multipath is the norm. Consequently, Doppler-based localization algorithms must trade rapid frequency dynamics against long observation windows while remaining resilient to unknown sound-speed profiles. The four contributions summarized below map a clear progression from fundamental Doppler estimation under multipath, through joint delay – Doppler exploitation, to bias-aware optimization of multistate geometries.

- **Characterizing Doppler estimation under multipath.** The starting point is a meticulous error analysis of a conventional DFT estimator in narrow-band acoustic links [55]. Analytic and simulation results show that the mean-square error and CRB decay with the record length quadratically, implying that a few-second snapshot suffices to average out most multipath corruption. This insight justifies longer coherent integration in later, delay-aided algorithms.
- **Trajectory-aware bistatic localization for the Internet of Underwater Things.** In a network of autonomous nodes, Zhang et al. formulate localization as a coupled design problem: choose observer trajectories that maximize the Fisher information derived from both time-of-arrival and Doppler cues [45]. By embedding a CRLB-based cost into an augmented Lagrangian optimizer, the method balances vehicle kinematics against obstacle avoidance. It raises positioning accuracy well beyond TOA only baselines – an illustrative example of geometry-control rather than post-processing cure.
- **Multistate tracking of maneuvering targets.** Differential delays alone are insufficient when both transmitters and receivers form a sparse multistate constellation; adding Doppler closes observability and tightens the CRLB, especially at short ranges. A two-stage algebraic estimator first linearizes the delay-Doppler equations via auxiliary variables, then applies a bias-correcting refinement that dispenses with iterative searches yet reaches near-bound performance at low SNR [46].
- **Bias-reduced semidefinite programming for uncertain sound speed.** Building on the previous idea, a more recent study recasts differential delay and Doppler measurements into a pseudolinear form amenable to WLS, which is then relaxed to a convex SDP with second-order constraints [47]. Because the sound velocity may fluctuate with depth and temperature, the solver explicitly models that uncertainty and adds a post-hoc bias correction. Simulations demonstrate mean-square errors almost indistinguishable from the theoretical bound across a wide speed-of-sound mismatch range.

Collectively, these works demonstrate that underwater Doppler localization has evolved from single-link frequency-tracking [55] to network-level, optimization-centric frameworks that exploit the full delay–Doppler manifold [45], [46], [47]. Key research includes: long coherent windows mitigate multipath; joint design of vehicle paths and estimators is vital in IoUT; and convex relaxations augmented by bias compensation can deliver bound-tight performance even when environmental parameters are poorly known.

### 4.3 *Tunnels, confined transport corridors and other harsh propagation environments*

Positioning systems that rely on satellite signals quickly degrade in shielded infrastructures such as road or rail tunnels, underground depots, or canyon-like industrial sites. Severe multipath, sparse anchor geometry and the impossibility of time-of-flight calibration demand hybrid schemes that fuse Doppler dynamics with complementary observables and robust state-space filtering. The three studies reviewed below illustrate how these constraints are tackled across radio, acoustic and vibro-acoustic modalities.

- **LTE-V Doppler/ToA fusion for road tunnels.** Halili et al. retrofit an existing vehicle-to-infrastructure (V2I) deployment with an extended Kalman filter that jointly tracks Doppler shifts and ToA from roadside LTE-V base stations [43]. Real-world measurements in the Beveren tunnel (Belgium) show mean localization errors of 20 m without additional hardware, outperforming plain ToA filtering by a factor of two. The work demonstrates that frequency dynamics captured by inexpensive on-board radios can compensate for the GNSS outage windows encountered by connected-vehicle fleets.
- **Delay-Doppler trajectory optimization for underwater corridors.** Although developed for bistatic sonar networks in the IoUT, the optimization framework in [45] is conceptually identical to the tunnel case: observers must maneuver within narrow, obstacle-laden corridors where GNSS is unavailable. By embedding both time-delay and Doppler terms into a Cramér–Rao inspired A-optimality cost and solving it with an augmented-Lagrangian method, the authors obtain observer trajectories that maintain Fisher information while respecting no-go zones and kinematic limits. The resulting paths boost positioning accuracy by up to 40 % relative to ToA-only navigation, highlighting the value of explicitly planning for strong Doppler diversity.
- **Pass-by acoustic localization of railway noise sources.** In wayside rail monitoring, neither satellites nor radio anchors are available; localization must rely solely on sound recordings. The Doppler-based Position Estimation (Db-PE) method in [63] tracks the instantaneous frequency of a passing train's noise via a short-time Fourier transform and an adaptive group-delay (AGCD) extractor. A nonlinear least-squares Doppler correction yields source kinematics, after which WLS over a microphone array pinpoints the track-side trajectory with markedly higher robustness than conventional beamforming approaches.

Across very different physical layers – cellular radio, sonar and acoustic noise – successful localization in GNSS-denied corridors follows a recurring pattern: harvest Doppler information that would otherwise be discarded, fuse it with time or angle measurements in a recursive estimator or trajectory-planning loop, and design algorithms that tolerate sparse infrastructure and strong multipath. These principles lay the groundwork for resilient positioning services in future connected-vehicle, industrial and subterranean networks.

### 4.4 *Radar-centric and UAV-centric scenarios*

Radar platforms have long exploited Doppler measurements for range-rate estimation. Yet, recent research extends this capability toward full three-dimensional localization, classification and even joint communication-sensing in highly dynamic aerial scenes. The papers reviewed here cover three complementary thrusts: (I) through-obstacle radar imaging with refined instantaneous-frequency modelling, (II) agile phased-array processing for clutter-limited environments, and (III) airborne RF ecosystems in which UAVs act as both targets and sensing platforms.

**Through wall and foliage penetrating radar localization.**

Two independent studies demonstrate that accurate tracking behind obstacles hinges on precisely modeling the target's instantaneous Doppler trajectory. The extended Bézier/Hough framework in [64] fits a two-shape-parameter Bézier curve to the time-frequency ridge, resolving frequency ambiguities that defeat short-time Fourier transforms. A follow-up algorithm couples Bézier fitting with an error-corrected Hough transform, further sharpening range-Doppler images and suppressing aliasing artefacts [65]. Both methods achieve real-time complexity ($O$(n)) and outperform classical STFT-peak approaches, making them attractive for search-and-rescue or urban combat scenarios where line-of-sight is blocked.

**Space-time adaptive processing for weather and air-traffic clutter.**

Phased-array weather radars must discriminate slow meteorological returns from fast aircraft echoes under severe spectral overlap. Kim et al. evaluate full-digital and sub-array phased-array architectures combined with space-time adaptive processing (STAP) filters [66]. By jointly exploiting spatial and temporal degrees of freedom, the system depresses clutter. It improves Doppler-velocity resolution, validating

STAP as a cornerstone for future multifunction radars that will concurrently sense and communicate.

**Doppler physics in six degree of freedom (6D) UAV mobility.**

Traditional channel models consider only translational motion; however, UAVs also yaw, pitch and roll. The stochastic geometry model in [67] derives closed-form macro- and micro-Doppler spectra for UAV-to-vehicle links under full 6D mobility, revealing that rotational components significantly reshape the power spectral density and time-correlation functions. These insights are vital when designing Doppler-assisted localization or waveform-adaptation schemes for aerial relays.

**UAVs as localization sensors.**

Moving from being targets to sensors, a single-rotor UAV equipped with a SDR performs SDF localization of phase-modulated signals [68]. Flight trials show meter-level accuracy across BPSK, QPSK and 16QAM emitters, provided that Doppler-shift vectors are long enough and the air-frame maintains speed stability – pointing to practical implementation guidelines for airborne electronic-support missions.

**Radar detection and classification of small drones.**

Conversely, ground sensors must detect hostile micro-UAVs that exhibit signature faintness and agility. A bistatic radar pipeline that learns micro-Doppler signatures through an LSTM network [52] achieves simultaneous detection, multitype classification and AOA localization. LSTM outperforms Support Vector Machine (SVM) and Convolutional Neural Network (CNN) baselines, confirming the efficacy of sequence-aware deep learning in extracting periodic blade-rotation features even with inexpensive hardware.

**Toward 6G joint sensing and communication.**

All of the above advancements converge with the wider trend of ISAC. Wei et al. survey OFDM, OTFS and GFDM waveforms, clarifying how pilot structures can be reused to retrieve distance, angle and Doppler while still carrying user data [53]. Their road-map underscores the need for tight synchronization control, flexible frame layouts and machine-learning-ready datasets – topics echoed throughout the radar and UAV literature.

Radar-centric localization now extends far beyond classical range-Doppler plots: advanced curve-fitting and STAP lift performance through obstacles and clutter, six-degree-of-freedom channel models expose new Doppler observables for aerial actors, and deep networks translate micro-Doppler patterns into reliable drone intelligence. Equally important, UAVs emerge as agile sensing nodes, and 5G-A/6G research is crystallizing the waveform foundations needed to fuse these capabilities into a seamless RF landscape.

### 4.5 *Satellite and geostationary scenarios*

Although most Doppler-based localization research concentrates on terrestrial or airborne sensors, LEO satellites have exploited the frequency-rate signature of emergency beacons for decades. A seminal contribution in this domain demonstrated that Doppler tracking can be significantly enhanced when complemented by purely frequency-reference measurements from GEO platforms [69].

The COSPAS-SARSAT study in [69] shows that GEO assets – despite providing no range-rate diversity – supply highly accurate carrier-frequency estimates that tighten the solution space defined by the LEO Doppler curves. Fusing the two data types yields three principal benefits:

- **Observation reduction**. A beacon location that normally requires three or more LEO passes can often be resolved with only two Doppler contacts when a single GEO measurement fixes the absolute transmit frequency.
- **Latency mitigation**. In sparse-access scenarios (e.g., polar or oceanic regions) the hybrid scheme advances the first-fix time by several minutes, which is critical for search-and-rescue response.
- **Solution disambiguation**. Simulation and archival-data analysis indicate a 13 % decrease in unresolved cases, mainly because the GEO constraint eliminates mirror-image solutions along the LEO ground track.

Sensitivity analysis further reveals that a 1 Hz systematic bias in the GEO frequency estimate translates into a position error of only a few hundred meters – acceptable within COSPAS–SARSAT requirements and indicative of the robustness of the hybrid Doppler/frequency-reference concept.

Looking ahead, these findings foreshadow broader opportunities in next-generation satellite constellations. Multi-orbit architectures (LEO-MEO-GEO) can leverage wideband payloads and precise onboard oscillators to combine Doppler, TDOA and AOA cues, enabling global, low-latency geolocation of non-cooperative emitters, IoT trackers and distress beacons. The GEO-assisted strategy outlined in [69] thus remains a touchstone for contemporary satellite-based Doppler localization research.

## 5 MEASUREMENT PLATFORMS AND FIELD IMPLEMENTATIONS

The previous sections have shown that Doppler-based localization theory now spans closed-form algebra, Bayesian inference and deep learning. Turning those algorithms into operational systems exposes a very different set of constraints: oscillator stability, processing latency, payload size, power budget and increasingly airworthiness of small UAV. This chapter surveys nine representative prototypes and testbeds, highlighting how hardware choices and experimental methodology shape achievable accuracy, reliability and cost. The case studies range from laser velocimeters mounted on micro-UAVs through low-cost SDRs to cooperative swarms and GNSS-free landing aids. Together they trace a clear evolution from bench-top emulation to real-world, multi-agent deployments.

A first line of research explores non-RF Doppler sensors as stand-alone navigators. An airborne two-dimensional laser Doppler velocimeter (LDV) delivers centimetric dead-reckoning on a quad-rotor even when GNSS is denied; by combining orthogonal laser beams with a lightweight IMU, the platform limits pitch-induced bias and achieves a 0.48 % distance error over fully autonomous flights [70]. Extending the concept to three axes and Kalman-filter fusion is flagged as the logical next step toward sub-decimeter odometry.

However, the bulk of recent activity leverages commodity SDRs to implement SDF geolocation. A seminal prototype based on a USRP B200mini demonstrates real-time frequency tracking and graphical localization in LabVIEW/MATLAB, reaching single-digit-meter accuracy during urban drive tests [71]. To decouple algorithmic evaluation from flight logistics, a laboratory emulator drives the same radio with synthetic Doppler waveforms that mimic 1 km and 10 km engagements; despite an external atomic clock, residual oscillator wander still inflates the error to tens or even hundreds of meters, underscoring the need for hardware-in-the-loop calibration [12]. Complementary acoustic emulation of a tactical transceiver shows that audio-frequency surrogates can reproduce RF Doppler dynamics with 20-47 m error and thus offer a low-cost pathway for rapid sensor testing [16].

Oscillator quality emerges as a dominant limitation. A systematic survey of six low-cost SDRs, bench-marked with internal and rubidium references, quantifies how frequency instability propagates into kilometer-scale localization bias; only units paired with a chip-scale atomic clock reduce the median error below 10 m and satisfy the weight and power budget of battery-powered UAVs [13]. These findings directly inform the design of Autonomous System of Location radio EmitteRs (ASLER). This SDF sensor combines a CSAC-stabilized USRP B200mini, Raspberry Pi compute node and offline topographic maps. Field flights over open terrain report an average radial error of 8.7 m at 250 m standoff, validating the end-to-end performance of a fully integrated, open-architecture system [14].

Scaling up from a single drone to a cooperative swarm multiplies coverage and resilience. Monte-Carlo simulations of up to ten UAVs, each running an extended SDF algorithm, reveal how emitter speed, flight altitude and LOS/NLOS transitions interact; a novel effectiveness factor shows that judicious swarm sizing can double track continuity in dense urban streets [1]. Beyond surveillance, Doppler cues can even enable mission-critical flight operations: an automatic landing procedure uses narrowband beacons to guide a VTOL UAV below one-meter touchdown error on offshore platforms or mountain pads, providing a GNSS-free fallback for emergency recoveries [17].

Finally, the practical impact of multipath-aware Doppler estimators is assessed in a MATLAB channel emulator representing realistic Power-Azimuth spectra. Among three candidate frequency estimators, a maximum-weight variant slashes urban localization error from tens of meters to the single-digit range, confirming that estimator selection is as important as hardware quality in harsh propagation [9].

These prototypes demonstrate that accurate Doppler-based localization is no longer confined to high-end military receivers. By marrying carefully chosen oscillators with open-source SDRs, lightweight processors and, increasingly, laser or vision-grade inertial sensors, researchers have pushed cost, size and energy barriers low enough for routine deployment on small UAVs and mobile robots. Equally important, controlled emulation frameworks allow repeatable stress-testing of sensors under extreme Doppler dynamics before the first outdoor flight – accelerating the transition from theoretical promise to field-ready capability.

## 6 CHALLENGES AND OUTSTANDING QUESTIONS

Doppler-based localization has matured from laboratory proof-of-concepts to field trials, yet three systemic obstacles still separate research demonstrators from robust, scalable deployments. First, real-world propagation rarely offers a clean, LOS tone: NLOS paths and dense multipath bias frequency estimates, erase periodic structure and couple the unknown geometry with clock offsets. Second, commodity hardware drifts in carrier frequency, suffers from limited dynamic range and cannot always sustain the computational load of high-fidelity estimators. Third, many envisioned applications – autonomous driving, drone swarms, ad-hoc IoT networks – evolve in highly dynamic scenes where sensors and targets move, invalidating static-geometry assumptions and tightening latency budgets. The papers reviewed in this section illustrate how these challenges are being exposed and partially mitigated while revealing open technical gaps.

Propagation induced bias is most acute when only a subset of multipath components carries useful geometric information. A recent single-antenna study shows that, in 5G/6G networks, Doppler signatures of multiple NLOS paths can be exploited to solve both localization and radio-environment mapping without any time or phase synchronization between user equipment and the serving base station [51]. Although the proposed one-dimensional search reaches the CRB at moderate SNR, the authors also report a systematic degradation of velocity accuracy as user speed increases, underscoring a fundamental trade-off between spatial diversity and dynamic-range limits in oscillator tracking.

Clock asynchrony and hardware jitter appear again in sequential localization-and-synchronization of time-division broadcast systems. The LAS-SDT framework fuses Doppler shifts with TOA observations to co-estimate position, velocity and clock bias of a mobile node whose oscillator is free-running [19]. Simulation results demonstrate clear accuracy gains over TOA-only filters. Yet, the method assumes that Doppler can be extracted with kilohertz precision from low-cost radios – an assumption that remains to be validated under the strong temperature and vibration drifts typical of embedded devices.

Even when LOS dominates, idealized radar models break down if Doppler-frequency migration is not properly compensated. Doppler-Compensated Localization (DC-Loc) tackles this issue for automotive millimeter-wave sensors by introducing an explicit velocity-dependent correction term during radar-sub-map alignment [72]. Benchmarks on Car Learning Act (CARLA) show a 25 % reduction in translational error relative to state-of-the-art SLAM baselines, but the solution relies on high-rate radar odometry that may overload resource-constrained edge processors. How to down-sample frequency updates without losing the observability required for sub-lane accuracy remains open.

Dynamic scenarios where both the target and the sensor array are in motion pose an additional layer of complexity. A thorough Cramér–Rao analysis proves that neglecting sensor kinematics in joint delay – Doppler measurements can inflate the positioning error by an order of magnitude, particularly in slow-propagation channels such as underwater acoustics [48]. The hybrid Gauss-Newton and quasi-Newton solvers proposed by the authors partly alleviate this sensitivity, yet their convergence hinges on a good initial guess – a non-trivial requirement when only noisy Doppler cues are available.

Across all four studies, two unifying research questions emerge. How can Doppler observables be decoupled from oscillator instability and environmental bias without resorting to costly atomic clocks? And how can estimator complexity be scaled down to fit the real-time, low-power constraints of edge platforms mounted on cars, drones or sensor buoys? Addressing these problems will likely require joint progress in adaptive signal processing, hardware-software co-design and data-driven calibration. Until then, Doppler localization will remain a promising yet fragile technique whose ultimate performance is set as much by propagation physics and hardware economics as by algorithmic ingenuity.

## 7 . FUTURE RESEARCH DIRECTIONS

The steady advance of Doppler-centric localization has exposed an increasingly diverse research agenda that blends signal processing, artificial intelligence and networked robotics. Four broad themes dominate current discussions: (I) the infusion of machine-learning models that can recognize or even predict Doppler signatures; (II) truly three-dimensional, time-varying estimation where both sensors and targets maneuver; (III) large-scale, distributed deployments in the IoT/IoUT; and (IV) trajectory-aware mission planning that treats the motion of drones or autonomous platforms as an optimization variable rather than a disturbance. The papers reviewed below exemplify these trends and highlight unresolved issues that set the stage for the next wave of investigations.

AI tools already underpin several prototypes. Turbo Iterative Positioning uses belief-propagation and gradient refinement to fuse Doppler shifts with delay grids in OTFS waveforms, delivering near-Cramér–Rao accuracy for swarming UAVs without any clock synchronization [73]. Although the algorithm still relies on handcrafted message-passing schedules, its success indicates that probabilistic graphical models and neural surrogates can replace exhaustive search in high-dimensional Doppler–delay spaces. A complementary survey of SDF applications argues that deep networks will soon automate multipath identification, emitter association and spectrum management across electronic-reconnaissance and public-safety operations, enabling single-platform receivers to track dozens of simultaneous signals in cluttered spectra [6].

Three-dimensional, temporally adaptive localization emerges as both an opportunity and a challenge. In underwater networks where the propagation speed is eighty-thousand times slower than radio, Zhang et al. show that path planning itself becomes a critical design knob. By embedding delay-and-Doppler Fisher information into an augmented-Lagrangian optimizer, autonomous nodes can choose trajectories that maximize geometric diversity and halve the error floor observed with static paths [45]. Similar motion-aware strategies are beginning to appear in terrestrial radar SLAM and 5G/6G user-equipment positioning. However, achieving real-time performance with limited on-board compute remains an open question.

Large-scale, distributed measurement systems raise fresh scalability issues. Tahat et al. catalogue state-of-the-art algorithms for moving receivers and emphasize that FDOA and FOA methods deteriorate under NLOS unless complemented by particle filters, extended Kalman loops or DPD engines [21] [23]. The review also points to a research gap in joint calibration of clocks, carrier frequencies and antenna arrays within ad-hoc IoT networks. In this area, federated learning or self-supervised calibration could play an influential role. In the underwater realm, the bistatic IoUT study cited above [45] reveals how acoustic networks must account for platform kinematics, local sound-speed profiles and energy budgets simultaneously, suggesting that cross-layer optimization will be indispensable for practical deployments.

Finally, trajectory planning is no longer confined to robotics; it is becoming a native localization component. Optimizing the motion of sensors (drones, vehicles, buoys) can convert weak Doppler gradients into highly informative measurements, but only if planners respect flight dynamics, collision avoidance and communication constraints. Reinforcement-learning-based planners. Differentiable Doppler simulators are promising yet largely unexplored pathways toward real-time, mission-aware geolocation.

In summary, the community pivots from isolated signal-processing blocks toward holistic, learning-enabled ecosystems that jointly shape waveforms, sensor trajectories and network protocols around Doppler observables. Bridging the gap between analytic guarantees and data-driven adaptability – while keeping power consumption, latency and privacy within acceptable limits – remains the central open problem for the coming decade.

## 8 SUMMARY

The literature on Doppler-aided positioning now spans a spectrum of geometries, signal models and hardware budgets. We therefore group the state of the art into five logical families, each built around a distinct measurement concept yet sharing common modelling assumptions and estimator templates.

**(1) Single-receiver SDF methods.**

By correlating the instantaneous Doppler trace collected along a mobile path with the kinematic state of the platform, a lone sensor can recover two- or three-dimensional emitter coordinates. Foundational wave-equation solutions established the forward model [2], while the UAV test-bed demonstrated meter-level accuracy [3]. Extensions cover phase-modulated signals [4] and broad application surveys [6]. Successive work improves robustness, overlapping-window filtering [7], multi-emitter marker extraction [8], multipath-weighted frequency estimators [9], and derivative-augmented solvers that approach the CRB [10]. Prototype studies on USRP hardware validate 2-5 m errors in drive tests [71], quantify oscillator bias [12], benchmark six low-cost SDRs [13], and field the autonomous ASLER multirotor (8.7 m mean error) [14] with GNSS fusion [15]. Acoustic surrogates [16], VTOL landing aids [17] and cooperative UAV swarms [1] enlarge the deployment envelope, while pseudo-linear and hybrid TOA/DFS filters target real-time tracking [19] [40].

**(2) Multi-receiver Frequency-Difference-of-Arrival techniques.**

Differential Doppler measurements across a spatial baseline generate iso-Doppler hyperboloids whose intersections locate the target. The canonical model and CRLB analysis can be traced to [20]. Classical solvers alternate between grid-search ML and algebraic/convex relaxations [21]. Recent variants exploit sequential single-sensor differencing [22] and derivative-enhanced CFS/SDP solvers that lift blind cones [23]. Key research threads address oscillator stability, multipath bias and low-Doppler regimes – as summarized in [21] – and advocate hybridization with TOA or SDF to stabilize geometry.

**(3) DPD with Doppler cues.**

DPD bypasses intermediate TDOA/AOA extraction and evaluates a global likelihood over raw snapshots. Complexity-reduced ML via MEM–DPD [25], FFT reuse for OFDM [26], and particle-filtered Doppler–delay fusion [58] now offer near-optimal accuracy in real time. Multi-source and high-resolution variants leverage MVDR spectra [28], beamspace compression [30], MUSIC fusion [30], GPU-parallelized importance sampling [31] and covariance-level AOA–Doppler coupling [32]. Robustness to multipath is achieved through compressive LOS selection on massive arrays [34], joint spatio-temporal suppression [35], and off-grid sparse Bayesian updates [36]. Quantized and distributed implementations retain performance with one-bit samples [37] [47], low-rate statistics [38] or unsynchronized OFDM anchors [39].

**(4) Hybrid Doppler + delay/angle/range schemes.**

Fusing complementary observables breaks geometric degeneracies and sustains accuracy when Doppler diversity is weak. Closed-form PLE/BCPLE/WIV fusions of AOA + Doppler reach CRLB performance without iteration [40]. Two-stage WLS blends TDOA, FDOA and dFDOA [41], while altitude-constrained solvers accommodate sensor-motion errors [42] [84]. Recursive LAS-SDT filtering unites Doppler, TOA and clock-bias estimation [74]. Cooperative range-and-velocity networks attain centimeter-level precision via online Gauss–Newton refinement [44]. Bistatic sonar/radar hybrids add differential Doppler to delays for bound-tight localization [46], [47], [48]. Application-driven hybrids span Wi-Fi search-and-rescue [49] and comprehensive 5G/6G tutorials [21].

**(5) Cutting-edge extensions: derivatives, micro-Doppler and learning.**

Time-differentiated Doppler observables (DD) break traditional blind directions and remove grid searches [10]. Siamese-network channel charting embeds Doppler-induced phase shifts for global mapping with only four unsynchronized antennas [50], while single-antenna OTFS users localize multipath reflectors through Doppler fingerprints alone [51]. Micro-Doppler LSTM pipelines classify and track small UAVs on low-cost bistatic radar [52]. Surveys of ISAC waveforms outline how OFDM, OTFS and GFDM can jointly convey data and high-resolution Doppler information [53].

Collectively, these five families reveal a progressive enrichment of the measurement space (from frequency to derivatives), estimator sophistication (from algebraic closed-form to GPU-parallel ML and neural inference) and platform diversity (from manned aircraft to SDR-equipped micro-drones and underwater nodes). The remainder of this section analyses each category in detail, contrasting accuracy, computational load and resilience to oscillator drift, multipath and asynchronous operation.

**Strategic relevance for future networks.**

Doppler-aware localization is advancing across multiple platforms, each leveraging tailored signal models and processing strategies. In 5G/6G edge infrastructure, Doppler-resolving DPD and hybrid FDOA/TDOA engines can be embedded as software slices, offering network-native localization without additional hardware; micro-Doppler–enhanced ISAC waveforms further promise centimeter-level tracking of autonomous vehicles within a single frame. UAV swarms benefit from swarm-wise SDF and sequential FDOA techniques, enabling each drone to self-localize and police spectrum in GNSS-denied environments [1] [68]. At the same time, six-degree-of-freedom Doppler channel models guide adaptive waveform design for aerial relays [67]. In autonomous road traffic, Doppler-corrected radar maps improve millimeter-wave SLAM performance [75], and LTE/NR-V2X tunnels already fuse Doppler and TOA via EKF to achieve lane-level accuracy [43]. Finally, in satellite IoT and distress beacon scenarios, GEO-assisted LEO Doppler tracking reduces first-fix latency, paving the way for multi-orbit constellations that integrate Doppler, delay, and angle cues at global scale.

In conclusion, Doppler observables – augmented by derivatives, fused with auxiliary cues and interpreted through learning-aware estimators – are poised to become a first-class positioning primitive for the sensor-rich, autonomy-driven landscape of 6G and beyond. Continued progress hinges on oscillator stabilization, complexity-balanced algorithms and cross-layer waveform co-design. Still, the path from theory to field-ready capability is already well mapped by the prototypes and trials reviewed here.

Table 1. synthesizes the characteristics of the main families of methods discussed in this review,

comparing their requirements, advantages, limitations, and performance levels reported in the literature.

Table 1. Comparison of Main Families of Doppler-Based Localization Methods

| Method Family | Core Principle | Hardware / Infrastructure Requirements | Typical Application Environments | NLOS / Multipath Robustness | Reported Accuracy (Literature) | Advantages | Limitations |
|---|---|---|---|---|---|---|---|
| **SDF – Single-receiver Signal Doppler Frequency** | Analyzes the variation of the received carrier frequency over time during receiver motion; fits radial velocity trajectories to a geometric model. | 1 mobile receiver (SDR, UAV, vehicle), motion sensors, stable oscillator (CSAC or OCXO recommended). | Urban, rural, UAV missions, emergency response, VTOL landing in GNSS-denied areas. | Medium – improved with weighted filtering, fusion with GNSS/TOA/AOA, or offline maps. | 2–10 m in UAV/SDR field tests; down to ~1% relative error in acoustics. | Low cost, no need for synchronization, rapid deployment. | Sensitive to oscillator stability, dependent on motion geometry. |
| **FDOA – Multi-receiver Frequency Difference of Arrival** | Differential Doppler measurements between receiver pairs; position derived from intersection of iso-Doppler hyperboloids. | ≥ 2 synchronized receivers (or sequential single), precise clocks. | Terrestrial networks, ISR, IoT, spectrum monitoring. | Medium–high (when fused with TOA/AOA); sensitive to low Doppler gradients. | <10 m with good geometry and stable clocks; degraded in low Doppler. | Works without receiver motion (if source moves), high network accuracy. | Requires stable frequency reference, sensitive to NLOS. |
| **DPD – Direct Position Determination with Doppler** | Directly estimates position from raw signal samples by matching a cost function that includes Doppler. | Multi-antenna or multi-sensor stations, high data throughput to central processor. | Radar, cellular networks, ISR, multi-source environments. | High (with massive MIMO and LOS selection); robust under low SNR. | Often achieves CRB; <1 m in massive MIMO setups. | Handles multiple sources, no need for intermediate parameter estimation. | High computational and transmission demands; requires many antennas or wide bandwidth. |
| **Hybrid Doppler + TOA/AOA/TDOA** | Combines frequency information with time or angle measurements in sequential or algebraic estimators. | Diverse sensor set; usually requires time or angle measurements. | Urban NLOS, tunnels, IoUT, cooperative networks. | High – fusion reduces blind zones and improves stability. | cm–m level with good geometry and calibration; ~20 m in LTE-V tunnels. | Robust to low Doppler diversity, improves solution stability. | Requires extra measurement types and data integration. |
| **Advanced – Doppler derivatives, micro-Doppler, ML-based** | Uses time derivatives of Doppler, Doppler differences, micro-Doppler signatures, and ML algorithms (Siamese nets, LSTM) to improve observability. | May operate with few antennas and no synchronization; requires large training datasets for ML. | ISAC 5G/6G, UAV tracking, channel mapping. | Variable – ML can mitigate NLOS but needs environment-specific training. | Potentially <1 m; can also classify objects. | Richer geometric information, can work without synchronization. | Low industrial maturity, depends on training data quality. |

ACKNOWLEDGEMENTS

This work was developed within the framework of the research grants no. UGB/531-000059-W300-22/2025 on "Transmission properties of radio wave propagation environments in military applications" and no. UGB/531-000128-W300-22/2026 on "Possibility analysis of using modern technologies in communication systems and electronic warfare", sponsored by the Military University of Technology (WAT), Poland.